\documentclass[aps,prd,showkeys,showpacs,amssymb,
amsfonts,epsf,preprintnumbers,nofootinbib,superscriptaddress]{revtex4}

\usepackage[dvips]{graphicx}
\usepackage{bm,latexsym,amsmath,amssymb,amsfonts,color}

\def\be {\begin{equation}}
\def\ee {\end{equation}}
\def\ba {\begin{eqnarray}}
\def\ea {\end{eqnarray}}

\def\bi {\begin{itemize}}
\def\ei {\end{itemize}}
\newcommand\beq{\begin{eqnarray}}
\newcommand\eeq{\end{eqnarray}}
\newcommand{\bea}{\begin{eqnarray}}
\newcommand{\eea}{\end{eqnarray}}

\usepackage{graphicx}
\usepackage{dcolumn}
\usepackage{bm,morefloats,color}

\usepackage{amssymb,bm}

\def\X5sp{{\rm X}_5}
\def\Y3sp{{\rm Y}_3}
\def\Z3sp{{\rm Z}_3}

\begin{document}

\title{Degravitation features in the cascading gravity model}

\author{Parvin Moyassari}
\affiliation{Arnold Sommerfeld Center, Ludwig-Maximilian University,
Theresienstr.~37, 80333 Muenchen, Germany}
\author{Masato Minamitsuji}
\affiliation{Yukawa Institute for Theoretical Physics, Kyoto University,
Kitashirakawa-Oiwake Cho, Sakyo-ku, 606-8502 Kyoto, Japan}
\affiliation{Multidisciplinary Center for Astrophysics (CENTRA), Instituto Superior T\'ecnico, Lisbon 1049-001, Portugal.}

\begin{abstract}
We obtain the effective gravitational equations
on the codimension-2 and codimension-1 branes in the cascading gravity model.
We then apply our formulation to the cosmological case
and 
obtain the effective Friedmann equations on the codimension-2 brane,
which are generically 
given in terms of integro-differential equations.
Adopting an approximation for which the thickness of the codimension-2 brane 
is much smaller than the 
Hubble horizon,
we study the Minkowski
and de Sitter codimension-2 brane solutions.
Studying the cosmological solutions shows that the cascading model exhibits the features necessary for degravitation of the cosmological constant. 
We also show that 
only the branch 
which does not have the smooth limit to 
the self-accelerating branch in five-dimensional model
in the absence of the bulk gravity
can satisfy the null energy condition
as the criterion of the stability.
Note that 
our solutions 
are obtained 
in 
a different setup
from that of the original
cascading gravity model
in the sense 
that the codimension-1 brane
contains matter fields other than the pure tension.
\end{abstract}
\maketitle

\section{Introduction}
The cosmological constant problem is one of the most pressing conceptual problems in physics.
This problem arises because the observed value of the vacuum energy is very small as compared to the values inferred from quantum field theory. 
Recently, a braneworld model 
which could provide a promising framework for addressing the cosmological constant problem has been developed.
This model is the so-called
cascading gravity model \cite{6,7,7a,gfree2}, which could induce infrared modifications of gravity that can screen the effect of a cosmological constant. This idea, referred to as ``degravitation'' \cite{6,degra},
could possibly provide a dynamical solution to the cosmological constant problem, 
since
any large cosmological constant initially present would degravitate away over time.

In the cascading gravity model,  which is a generalization of the 
Dvali-Gabadadze-Porrati (DGP) model \cite{3} to higher dimensions,
one constructs a sequence of
branes with decreasing dimensions placed
on each one,
where each
brane action contains the induced gravity term. The gravitational force falls off faster 
at large distances in the cascading model than in
the original five-dimensional (5D) DGP model \cite{7}.
In the simplest six-dimensional (6D) cascading model,
our four-dimensional (4D) universe, codimension-2 brane,
is placed on a codimension-1 brane
which is embedded into a 6D bulk spacetime.
The 6D model contains two crossover scales,
$r_3:=\frac{M_4^2}{M_5^3}$
and 
$r_4:=\frac{M_5^3}{M_6^4}$,
where $M_6$, $M_5$ and $M_4$ are gravitational
energy scales in the bulk and on the codimension-1 and codimension-2 branes,
respectively.
Assuming that $r_3\ll r_4$,
it is expected that
the gravitational potential on the codimension-2 brane cascades from 
the 4D regime at short scales, 
to the 5D one at intermediate distances
and finally to the 6D regime at large distances.
This model addresses a problem in the 6D
brane world models with the induced gravity.
If there is no induced gravity term on the codimension-1 brane, 
the bulk graviton propagator diverges logarithmically
near the position of the codimension-2 brane. 
Then the energy scale $r_4^{-1}$ 
acts as an infrared cutoff for the propagator \cite{gfree2}
so that it remains finite even at the position of the codimension-2 brane.
The cascading model suffers a ghost instability
if there is no tension on the codimension-2 brane.
However, very interestingly, it has been shown that
there is a critical tension
above which the model 
becomes ghost-free \cite{gfree2,gfree}. 
This model may also provide a mechanism for
the degravitation \cite{7,7a}
which can support a very small expansion rate 
of our universe even in the presence of a large
cosmological constant, namely the codimension-2 brane tension.
Hence, the degravitation could provide a way 
to resolve the cosmological constant problem.

The purpose of this paper
is to see 
whether in reality
there could be some features of degravitation
in the cascading gravity model.
 In order to establish if this model exhibits degravitation it is necessary to understand its cosmological evolution and obtain the effective Friedmann equations. The cosmological behavior can
be very different from the ordinary 4D cosmology
and the 5D DGP model,
although they should be recovered in a certain limit.
Cosmology in the cascading gravity model has been 
studied in the context
of the 5D theory in \cite{cosmo},
which is composed of 
gravity coupled to a scalar field \cite{gal}, 
originated from the bending of the 
codimension-1 brane in the 6D bulk.
This theory 
is a nonlinear extension of 
the weak gravity limit of
the full 6D model,
but it is not the unique extension. Although the 5D theory may 
possess several similarities to the full 6D theory,
the final confirmation should be made
in the context of the original 6D theory.
A regular solution 
involving a flat codimension-2 brane with a tension 
has also been discussed in \cite{aga}.
Some implications of the cascading gravity model to cosmological
events have been discussed in, e.g., Refs. \cite{nb,mk}.
 
In this paper, we present the covariant formulation of the nonlinear effective gravitational theory on the boundaries, codimension-1 and codimension-2 branes.
We then apply them to find the cosmological solutions,
in particular, the de Sitter solutions 
which may possess 
features of 
degravitation.
In the gravitational equations on the codimension-2 brane,
the bulk contributions are
given in terms of the integration over the sixth direction
where we take the finite thickness of the codimension-2 brane
into consideration.
We study some cosmological solutions that show degravitation of the cosmological constant in the cascading model. 
In these solutions the effect of
the effective vacuum energy
on the Hubble expansion rate
is suppressed
by the ratio $\frac{r_3}{r_4}$
in comparison with 
the naive expectation
from the ordinary 4D cosmology.
Moreover, we will discuss some self-accelerating solutions
by applying the small thickness approximations
where the codimension-2 brane's thickness is
assumed to be much smaller than the size of the Hubble horizon.
We also show that 
only the branch 
which does not have the smooth limit to 
the self-accelerating branch in the 5D model
in the absence of the bulk gravity
can satisfy the null energy condition
on the codimension-1 brane.
Note that 
our solutions 
are obtained 
in a different setup
than the original
cascading gravity model \cite{6,gfree2}
in the sense
that the codimension-1 brane
contains matter fields other than the pure tension.

\section{Gravitational equations in the cascading gravity model} 

In the simplest realization of the cascading gravity our 
codimension-2 brane world
is placed on a codimension-1 brane embedded into a 6D bulk spacetime. 
We start from the general Arnowitt-Deser-Misner (ADM) form of the 6D metric,
with the bulk coordinate $y$ playing the role
of a time variable,
\footnote{The 6D metric is 
\beq \nonumber
g_{AB}= \Big(
\begin{array}{cc}
 N^2+ N^aN_a  & N_a \\
N_a  & g_{ab}   
\end{array}
\Big).
\eeq}
\beq
\label{3}
{}^{(6)}ds^2=N^2dy^2+g_{ab}(dx^a+N^a dy)(dx^b+N^b dy),
\eeq
where $g_{ab} $ is the 5D induced metric on the codimension-1 brane with $a,b=z,\mu$ (here Greek letters denote the 4D space indices.). 
Having the bulk and brane coordinate systems, 
the full 6D action can be written
as
\bea
\label{action}
S&=&\frac{M_6^4}{2}\int
d^6 x \sqrt{-{}^{(6)}g}~
{{}^{(6)}R}
 +\int
 d^6 x \sqrt{-{}^{(5)}g}
\Big(\frac{M_5^3}{2}~{{}^{(5)}R}+{\cal
L}_{5}^{mat}\Big)
\delta(y)
\nonumber\\
&+&
\int
 d^6 x 
\sqrt{-{}^{(4)}g}
\Big(\frac{M_4^2}{2}~{{}^{(4)}R}+{\cal
L}_{4}^{mat}\Big)
\delta(y)\delta(z).
\eea
\begin{figure}
 \begin{center}
\includegraphics[width=0.58\textwidth]{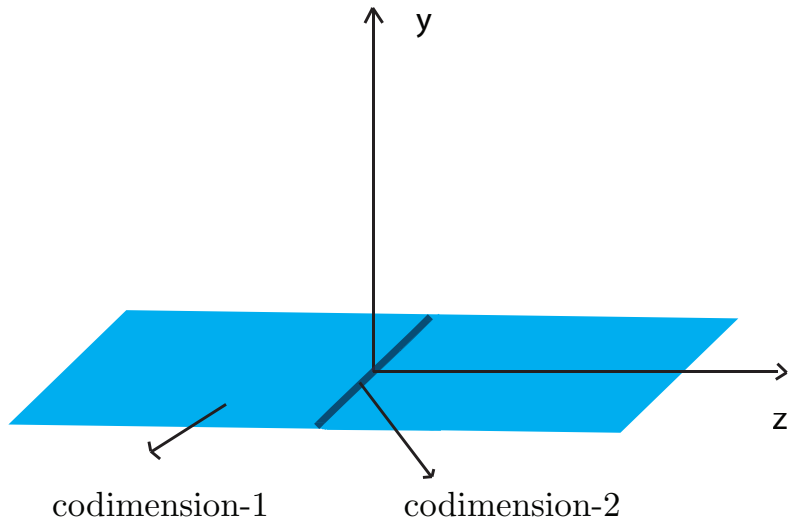}
\caption{
The configuration of the
codimension-1 (plane) and codimension-2 (thick line)
branes in the six-dimensional bulk is shown.
The induced metric on the codimension-1 brane is $g_{ab}$ and on the codimension-2 brane is $g_{\mu\nu}$.
}
\label{fig2}
\end{center}
\end{figure}
The codimension-1 brane is located at $y=0$,
 and the codimension-2 brane is placed on the codimension-1 brane 
at $y=z=0$ (see Fig. 1).
We are interested in the derivation of
the gravitational equations on the codimension-2
brane. 
In the ADM formalism, the full 6D action for the cascading setup, including the appropriate Gibbons-Hawking
 boundary term in the 6D part, can be rewritten as
\beq\label{5}
S
&=&\frac{M_6^4}{2}
\int 
d^6x
\sqrt{-{}^{(6)}g}~
({}^{(5)}R
+K^{2}
-K_{ab}K^{ab})
+\int d^6 x
\sqrt{-{}^{(5)}g}~
\Big(\frac{M_5^3}{2}{{}^{(5)}R}+{\cal
L}_{5}^{mat}\Big)\delta(y)\nonumber\\
&+&
\int d^6 x 
\sqrt{-{}^{(4)}g}~
\Big(\frac{M_4^2}{2}{{}^{(4)}R}+{\cal
L}_{4}^{mat}\Big)\delta(y)\delta(z),
\eeq
where the extrinsic curvature $K_{ab}$ is given by
 \beq\label{6}
 K_{ab}=\frac{1}{2N}(\partial_y g_{ab}-\nabla_aN_b-\nabla_b N_a),
 \eeq
and $K=K_{ab}g^{ab}$. 
To derive the gravitational equations we should take the variation of the action (\ref{5}) with respect to the variables $N$, $N_a$ and $g_{ab}$ which
vanish at infinity but 
are nonzero on the boundary branes. 
Thus 
the boundary terms 
arising in the variations
give the 
contributions to the gravitational theory 
on the hypersurface of $y=0$ (codimension-1 brane) 
and the hypersurface of $y=z=0$ (codimension-2 brane).
In the following sections,
we consider the variation of the 6D, 5D and 4D parts of the action 
and derive the 
effective gravitational theory
on the codimension-1 and codimension-2 branes,
separately.

\subsection{Variation of the 6D part of the action} 

Taking the variation of the bulk terms with respect to $N$, $N_a$ and $g_{ab}$ gives the bulk Einstein equations 
\beq\label{7}
{}^{(6)}G_{AB}=0, ~~~ ~A,B=y,z,\mu.
\eeq
The boundary contributions in the variation of the 6D part of the action 
give the contributions
localized
to the hypersurface $y = 0$. 
Since there is no term coming from $\delta\partial_y N$ and $\delta\partial_yN_{a}$ the only terms localized to the codimension-1 brane come from the term proportional to $\delta\partial_y g_{ab}$. Since $\delta g_{ab}$ does not vanish on the brane, 
the
boundary contributions to the variations of the 6D part of the action read
\beq\label{8}
-\frac{M_6^4}{2}\Delta_y\int  dzd^4x\sqrt{-{}^{(5)}g}~(Kg^{ab}-K^{ab})\delta g_{ab}.      
\eeq   
 Here we have used $\sqrt{-{}^{(6)}g}=N\sqrt{-{}^{(5)}g}$ and
$\triangle_y$ is the discontinuity in a given quantity $A$ over the codimension-1 brane namely,
$\triangle_y A=2A|_{y=0+}$.
From now on we omit the subscript $~\big|_{y=0+}$.

\subsection{Variation of the 5D part of the action} 

Now we take the variation of the 5D part of the action (\ref{5}) with respect to $g_{ab}$ which gives us the contribution of the 5D part on the codimension-1 brane. 
Since there is a codimension-2 brane embedded in the codimension-1 brane, 
there are also 
contributions localized to the hypersurface $y=z=0$.
To obtain the 5D contributions 
localized to the codimension-2 brane, 
we go to the ADM form of the 5D metric $g_{ab}$ adapted to the codimension-2 brane 
where $z$ plays the role of a time variable,
\footnote{The 5D metric is 
\beq  \nonumber
g_{ab}= \Big(
\begin{array}{cc}
  \mathcal{N}^2+ \mathcal{N}^\mu\mathcal{N}_\mu  &  \mathcal{N}_\mu \\
\mathcal{N}_\mu  & g_{\mu\nu}   
\end{array}
\Big).
\eeq}
\beq\label{13}
^{(5)}ds^2=g_{ab}dx^a dx^b
=
\mathcal{N}^2dz^2+g_{\mu\nu}(dx^\mu+\mathcal{N}^\mu dz)(dx^\nu+\mathcal{N}^\nu dz),
\eeq
where $g_{\mu\nu}$ is the 4D induced metric on the  codimension-2 brane.
In the ADM formalism in the 5D spacetime,
the 5D part of the action (\ref{action}) 
can be rewritten as
\beq\label{14}
\int d^6 x\sqrt{-{}^{(5)}g}
\Big(\frac{M_5^3}{2} ~{{}^{(5)}R}
+{\cal
L}_{5}^{mat}\Big)\delta(y)
&=&\frac{M_5^3}{2}\int  d^6x \sqrt{-{}^{(5)}g}({}^{(4)}R
+\mathcal{K}^{2}
-\mathcal{K}_{\mu\nu}\mathcal{K}^{\mu\nu})\delta(y)\nonumber \\&+&\int d^6x\sqrt{-{}^{(5)}g}{\cal
L}_{5}^{mat}\delta(y),
\eeq
where
\beq
\label{14a}
\mathcal{K}_{\mu\nu}
=\frac{1}{2\mathcal{N}}(\partial_z g_{\mu\nu}-\nabla_\mu \mathcal{N}_\nu-\nabla_\nu \mathcal{N}_\mu),
 \eeq
and $\mathcal{K}=g^{\mu\nu}\mathcal{K}_{\mu\nu}$. 
Varying the 5D part of the action with respect to $g_{ab}$
leads to 
\beq\label{15}
\frac{1}{2}\int dz d^4x\sqrt{-{}^{(5)}g}({M_5^3}~{}^{(5)}G_{ab}-S_{ab})\delta g^{ab},
\eeq
which is localized to the hypersurface $y=0$.
Here $S_{ab}$ is the energy momentum tensor for matter 
on the codimension-1 brane. The contributions localized to the codimension-2 brane 
come from the variation $\delta \partial_z  g_{\mu\nu}$.
Note that there is no term which is proportional to
$\delta \partial_z\mathcal{N}_{\mu}$ and $\delta\partial_z\mathcal{N}$ 
in the variation of the 5D part of the action.
Thus
the only boundary term localized to the codimension-2 brane
coming from the variations of the 5D part of the action is
\beq\label{15a} 
-\frac{M_5^3}{2}\triangle_z\int d^4x\sqrt{-{}^{(4)}g}(\mathcal{K}g^{\mu\nu}-\mathcal{K}^{\mu\nu})\delta g_{\mu\nu},
\eeq
where we have used $\sqrt{-{}^{(5)}g}=\mathcal{N}\sqrt{-{}^{(4)}g}$.  
The discontinuity in a given quantity $A$
across the hypersurface of $z=0$ is 
given by $\triangle_z A=2A|_{z=0+}$.
From now on we omit the subscript $|_{z=0+}$.

\subsection{Variation of the  4D part of the action}

Finally, 
varying the 4D part of the action with respect to $g_{ab}$ leads to
\beq\label{15b} 
\frac{1}{2}\int d^4x\sqrt{-{}^{(4)}g}({M_4^2}~{}^{(4)}G_{\mu\nu}-T_{\mu\nu})\delta^\mu_a\delta^\nu_b\delta g^{ab},
\eeq
where $T_{\mu\nu}$ is the energy momentum tensor for matter 
on the codimension-2 brane.

\subsection{Effective gravitational equations on the branes }
Collecting the boundary contributions 
in the variations of all parts of the action (\ref{action}) 
derived in the previous sections, 
one can obtain the effective gravitational equations on the 
codimension-1 and codimension-2 branes. 
Combining (\ref{8}) with (\ref{15}), (\ref{15a}) and (\ref{15b}),
the boundary equations on the 
codimension-1 brane at $y=0$
read
\bea\label{9}
\int dz d^4x \sqrt{-{}^{(5)}g}~(M_5^3 {}^{(5)}G_{ab}-S_{ab})&=&
\Delta_y
\int dz d^4x\sqrt{-{}^{(5)}g}~{M_6^4}(K_{ab}-Kg_{ab})
\\
&+&\int dz d^4x\sqrt{-{}^{(4)}g}~\Big[-
M_4^2{}^{(4)}G_{\mu\nu}+T_{\mu\nu}+2{M_5^3}(\mathcal{K}_{\mu\nu}-\mathcal{K}
g_{\mu\nu})\Big]\delta^\mu_a\delta^\nu_b\delta(z).\nonumber
\eea
Now we assume that the shift vector $N_a=n_a(x)
s(y)
\epsilon(z)$.
 The function 
$\epsilon(z)$
is a regulating function with the following properties: $\epsilon(\infty) = 1$, $\epsilon(-z) = -\epsilon(z)$ and $\epsilon(z)_{,z}= 2\delta_\epsilon(z)$, where $\delta_\epsilon(z)$ is a regularization of the Dirac delta function. 
The function $s(y)$ is the sign function,
and $s(y)_{,y}=2\delta (y)$.
Using this ansatz one can see that the first term on the right-hand side of  (\ref{9}) includes 
some terms localized to the codimension-2 brane. 
One can see this by  using the components of the extrinsic curvature $K_{ab}$,
\beq\label{11}
  K_{\mu\nu}&=&\frac{1}{2N}\Big[\partial_yg_{\mu\nu}-\partial_{(\mu} N_{\nu)}+2\Gamma_{\mu\nu}^a N_a \Big]
={\tilde K}_{\mu\nu},
\nonumber\\
   K_{zz}&=&
\frac{1}{2N}\Big[\partial_yg_{zz}-4n_z(x)
s(y)
\delta_\epsilon(z) +2\Gamma_{zz}^a N_a \Big]
={\tilde K}_{zz}
-\frac{2n_z(x)s(y)}{N}\delta_\epsilon(z)
,\\
 K_{z\mu}&=&\frac{1}{2N}\Big[\partial_yg_{z\mu}
-2n_\mu(x)
s(y)\delta_\epsilon(z)-\partial_\mu N_z+2\Gamma_{\mu z}^aN_a\Big]
={\tilde K}_{z\mu}
-\frac{n_\mu(x)s(y)}{N}\delta_\epsilon(z),
\nonumber
\eeq
where ${\tilde K}_{ab}$ represents the extrinsic curvature
except for the terms proportional to $\delta_\epsilon(z)$.
Thus the gravitational equations on the codimension-1 brane at $y=0$,
off the codimension-2 brane at $z=0$,
are given by
\bea
\label{11_1}
 {}^{(5)}G_{ab}=\frac{1}{M_5^3}S_{ab}
+\frac{2M_6^4}{M_5^3}({\tilde K}_{ab}-{\tilde K}g_{ab}).
\eea

Finally, let us derive the gravitational equations on the 
codimension-2 brane at $y=z=0$.
Plugging the components of (\ref{11}) into (\ref{9}), and taking the integral over $z$ across the codimension-2 brane,
we obtain the terms localized to the codimension-2 brane coming from the boundary contribution to the variations of the action (\ref{action}). 
Therefore, the boundary equations on the codimension-2 brane read
\beq\label{17}
 { n_\nu(x)=0, 
}
\eeq

\bea\label{18}
{
{}^{(4)}G_{\mu\nu}=\frac{1}{M_4^2 }T_{\mu\nu}+2\frac{M_5^3}{M_4^2 }(\mathcal{K}_{\mu\nu}-\mathcal{K}g_{\mu\nu})
+4\frac{M_6^4}{M_4^2}\int dz \frac{{\mathcal{N}}}{N}n_z(x)\Big(g^{zz}g_{\mu\nu}-g^{z}_{~\mu}g^{ z }_{~\nu}\Big)\delta_\epsilon(z)}.
\eea
Equations (\ref{9}), (\ref{17}) and (\ref{18}), 
along with the bulk Einstein equations (\ref{7}), 
are equations of motion in the cascading gravity model. It is obvious that when $M_6^4=0$, the DGP gravitational equations on the codimension-2 brane are recovered.


\section{Cosmology}

\subsection{
Restricting the bulk metric for a cosmological codimension-2 brane}

In this section, we apply the formulation developed in
the previous section to cosmology.

\subsubsection{
Inhomogenenous bulk metric}

First,
we present the form of the most general bulk metric 
where a general cosmological codimension-2 brane 
is embedded.
We assume the Friedmann-Robertson-Walker (FRW)
metric on the codimension-2 brane,
\bea
\label{flat_frw}
ds_4^2=-dt^2+a(t)^2 \delta_{ij}dx^i dx^j,
\eea
where $a(t)$ is the scale factor of the
flat three-dimensional space.

Starting from the most general bulk metric 
in the ADM form,
\bea
\label{hom_6d}
ds_6^2
&=&N(t,x^i,y,z)^2 dy^2
+g_{zz}(t,x^i,y,z)\big(dz+N^z(t,x^i,y,z)dy\big)^2
\nonumber\\
&+&g_{tt}(t,x^i,y,z)\big(dt+N^t(t,x^i,y,z) dy\big)^2
+g_{ij}(t,x^i,y,z)(dx^i+N^i(t,x^i,y,z) dy)(dx^j+N^j(t,x^i,y,z) dy)
\nonumber\\
&+&2g_{zt}(t,x^i,y,z)\big(dz+N^z(t,x^i,y,z)dy\big)
                     \big(dt+N^t(t,x^i,y,z)dy\big)
\nonumber\\
&+&2g_{zi}(t,x^i,y,z)(dz+N^z(t,x^i,y,z)dy)(dx^i+N^i(t,x^i,y,z) dy)
\nonumber\\
&+&2g_{ti}(t,x^i,y,z)
\big(dt+N^t(t,x^i,y,z)dy\big)\big(dx^i+N^i(t,x^i,y,z)dy\big),
\eea
along the codimension-1 brane, $y=0$,
(\ref{hom_6d}) reduces to
\bea
\label{hom_5d}
ds_5^2
&=&
 g_{zz}(t,x^i,0,z)dz^2
+g_{tt}(t,x^i,0,z) dt^2
+g_{ij}(t,x^i,0,z)dx^i dx^j
\nonumber\\
&+&2g_{zt}(t,x^i,0,z)dt dz
+2g_{zi}(t,x^i,0,z)dz dx^i
+2g_{ti}(t,x^i,0,z)dt dx^i,
\eea
while
along the codimension-2 brane, $y=z=0$,
(\ref{hom_5d}) reduces to
\bea
\label{hom_4d}
ds_4^2= 
 g_{tt}(t,x^i,0,0) dt^2
+g_{ij}(t,x^i,0,0)dx^i dx^j
+2g_{ti}(t,x^i,0,0)dt dx^i.
\eea
The metric (\ref{hom_4d}) is 
matched with the FRW metric (\ref{flat_frw}),
and 
\bea
g_{tt}(t,x^i,0,0)=-1,\quad
g_{ij}(t,x^i,0,0)=a(t)^2\delta_{ij},\quad
g_{ti}(t,x^i,0,0)=0.
\label{cosmo}
\eea
Under
our assumption on the shift vector $N_a=n_a(x)s(y)\epsilon(z)$
(see Sec. 2.4)
the constraints \eqref{17} require that
\bea
&&0=
n_{t}(t,x^i)s(y)\epsilon(z)
 = g_{zt}(t,x^i,y,z)N^z(t,x^i,y,z)
 +g_{tt}(t,x^i,y,z)N^t(t,x^i,y,z)
 +g_{tj}(t,x^i,y,z)N^j(t,x^i,y,z),
\nonumber\\
&&0=n_{i}(t,x^i)s(y)\epsilon(z)
 =g_{zi}(t,x^i,y,z)N^z(t,x^i,y,z)
 +g_{ti}(t,x^i,y,z)N^t(t,x^i,y,z)
 +g_{ij}(t,x^i,y,z)N^j(t,x^i,y,z).
\nonumber\\
&&
\label{constr}
\eea
Note that since $n_\mu(t,x^i)=0$,
$N_{t}$ and $N_{i}$ also become zero
in the whole bulk.
We also have the relation
\bea
\label{z-comp}
&&0
\neq 
n_z(t,x^i)s(y)\epsilon(z)
= g_{zz}(t,x^i,y,z)N^z(t,x^i,y,z)
 +g_{zt}(t,x^i,y,z)N^t(t,x^i,y,z)
 +g_{zj}(t,x^i,y,z)N^j(t,x^i,y,z).
\nonumber\\
&&
\eea
In (\ref{z-comp}), 
it is reasonable
to assume
$N^a(t,x^i,y,z)=n^a(t,x^i,y,z)s(y)\epsilon(z)$,
where $n^a(t,x^i,y,z)$ does not contain 
distributions,
and 
\bea
\label{z-comp2}
n_z(t,x^i)
= g_{zz}(t,x^i,y,z)n^z(t,x^i,y,z)
 +g_{zt}(t,x^i,y,z)n^t(t,x^i,y,z)
 +g_{zj}(t,x^i,y,z)n^j(t,x^i,y,z).
\eea
This constraint is severe,
in the sense that
the left-hand side of (\ref{z-comp2})
is independent of the position in the bulk 
while
the right-hand side depends on it.
Similarly,
the constraint equations \eqref{constr}
reduce to
\bea
\label{compcomp}
&&0=n_{t}(t,x^i)
 = g_{zt}(t,x^i,y,z)n^z(t,x^i,y,z)
  +g_{tt}(t,x^i,y,z)n^t(t,x^i,y,z)
  +g_{tj}(t,x^i,y,z)n^j(t,x^i,y,z),
\nonumber\\
&&0=n_{i}(t,x^i)
 =g_{zi}(t,x^i,y,z)n^z(t,x^i,y,z)
 +g_{ti}(t,x^i,y,z)n^t(t,x^i,y,z)
 +g_{ij}(t,x^i,y,z)n^j(t,x^i,y,z),
\eea
where 
all the components of $n^a(t,x^i,y,z)$ 
are assumed to be regular everywhere in the bulk.

Let us check the compatibility with
the boundary equations on the codimension-2 and
codimension-1 branes,
Eqs. (\ref{11_1}) and (\ref{18}).
Since the left-hand side of (\ref{18})
is homogeneous with respect to $x^i$,
the right-hand side should also be homogeneous.
The second term in the right-hand side of (\ref{18}),
with the definition (\ref{14a}),
depends on $\partial_z g_{\mu\nu}$.
All 
components of $\partial_z g_{\mu\nu}$
are independent of $x^i$ along 
$y=z=0$.
Thus along the codimension-2 brane, $y=z=0$,
$\partial_z{\tilde g}_{ij}$ ($\propto \delta_{ij}$)
and $\partial_z {\tilde g}_{tt}$
are independent of $x^i$,
and ${\tilde g}_{ti}=0$.

${\cal N}^{\mu}$ is written
as a combination of $g_{z\mu}$ and $g_{\mu\nu}$,
and through the definition of (\ref{14a})
it can induce an inhomogeneity in
the right-hand side of the first relation in (\ref{18}), 
if $g_{z\mu}$ is $x^{i}$ dependent along $z=0$.
Thus we may require that $g_{z\mu}$ is independent of $x^i$
along $z=0$.
Since ${\cal K}_{ij}\propto \delta_{ij}$,
we have to impose ${\cal N}_i=0$ 
and ${\cal N}_t= {\cal N}_t(t)$
along the codimension-2 brane,
which leads to 
${\cal N}_t=-{\cal N}^t (t)$
and 
${\cal N}_i=0$.
This also leads to 
$g_{zt}(t,x^i,0,0)=-{\cal N}^t(t)$
and 
$g_{zi}(t,x^i,0,0)=0$
on the codimension-2 brane.

\subsubsection{Homogeneous bulk metric}

With the homogeneity and isotropy on the 
codimension-2 brane,
we have imposed several boundary conditions
on the codimension-1 brane.
However, it is still quite difficult to 
solve the gravitational equations 
for 
an inhomogeneous metric ansatz.
Thus we 
make
further simplifications on the metric ansatz:
We impose homogeneity over the bulk and the branes,
and drop all the $x^i$ dependence
from the bulk metric (\ref{hom_6d});
$g_{AB}(t,x^i,y,z)\to g_{AB}(t,y,z)$.

Some 
additional simplification could also 
help to reduce the problem to a soluble one.
On the codimension-2 brane,
$g_{zi}$ and its first order derivative  
$\partial_z g_{zi}$ vanish.
For further simplicity, we 
assume that $g_{zi}(t,y,z)=0$ over the bulk.
Similarly,
we also 
assume that $g_{ti}(t,y,z)=0$ over the bulk.
Then, 
Eqs. (
\ref{compcomp}) reduce to 
\bea
&&0
 = g_{zt}(t,y,z)n^z(t,y,z)
  +g_{tt}(t,y,z)n^t(t,y,z),
\quad
0=
g_{ij}(t,y,z)n^j(t,y,z).
\label{const3}
\eea
From the second relation, $n^j(t,y,z)=0$.
Here 
we assume
that each term on the right-hand side of
Eq.  (\ref{const3}) is zero.
Since $g_{tt}(t,y,z)$ cannot be zero,
$n^t(t,y,z)=0$,
and also $n^z(t,y,z)=0$ or $g_{zt}(t,y,z)=0$.
If we set $n^z=0$, there is 
no effect on the codimension-2 brane from the bulk,
and we then assume
 $g_{zt}(t,y,z)=0$.
Taking Eq. (\ref{const3}) into account,
(
\ref{z-comp2}) becomes  $n_z(t)=g_{zz}(t,y,z)n^z(t,y,z)$.
Since the product is independent of the bulk coordinates,
$g_{zz}(t,y,z)= u(t,y,z) f_g(t)$
and $n^z(t,y,z)=u(t,y,z)^{-1}f_n(t)$,
where $u(t,y,z)$, $f_g(t)$ and $f_n(t)$
are smooth functions of the corresponding arguments.

From the above discussions,
the bulk metric finally reduces to 
\bea
\label{hom_6d_7}
ds_6^2
&=&N(t,y,z)^2 dy^2
+
f_g(t)u(t,y,z)
\big(
 dz
+\frac{f_n(t)}{u(t,y,z)}
  s(y)\epsilon(z)dy\big)^2
+g_{tt}(t,y,z)dt^2
+g_{ij}(t,y,z)dx^i dx^j.
\eea
For the given ansatz Eq. (\ref{hom_6d_7}),
the usual way to find a solution
is to solve the Einstein equations.
For the vacuum bulk,
the simplest nonsingular solution would be 6D Minkowski spacetime.
In the rest of the paper,
we will study the cosmological brane solutions
in the 6D Minkowski bulk.
In the 5D DGP model, 
the 
cosmological solution
was originally found
in the case of the 5D Minkowski bulk \cite{4,5}.
Similarly,
in 6D spacetime,
it is also reasonable to start from the case of the Minkowski bulk.
In the next subsection,
starting from 6D Minkowski spacetime,
we will construct the bulk metric
where 
the codimension-2 brane with a 4D FRW universe
is embedded.
We will show that
the 6D metric (\ref{hom_6d_7})
contains the case of Minkowski spacetime.


\subsection{The 6D Minkowski bulk}

We start from  the 6D Minkowski spacetime with the foliation
\bea
\label{com}
ds_6^2
&=&
(c^2-\beta^2)dy^2
+(dz+ \beta dy)^2
-\Big(1+\big(\frac{\beta}{c}z+cy\big)
\big(H+\frac{\dot{H}}{H}\big)
\Big)^2dt^2
\nonumber\\
&+&a(t)^2
\Big(1+\big(\frac{\beta}{c}z+cy\big)H\Big)^2
\delta_{ij}dx^i dx^j,
\eea
where $\beta$ and $c$ are constants,
$a(t)$ is an arbitrary function of $t$
and $H(t):=\frac{\dot{a}}{a}$.
The dot denotes the derivative with respect to $t$.
The metric (\ref{com}) clearly 
satisfies the 6D Einstein Eq. (\ref{7}).
Note that if $\beta$ and $c$ are time dependent
the metric 
is not a solution to Eq. (\ref{7}).
On the $y=z=0$ hypersurface
where the metric can be written as
\bea
ds_4^2=
-dt^2
+a(t)^2\delta_{ij}dx^i dx^j,
\eea
the function $a(t)$
can be interpreted as the cosmic scale factor
of the FRW universe,
where $H$ becomes the Hubble parameter.
Imposing $Z_2$ symmetry across the 
$y=0$ and $z=0$ hypersurfaces
and choosing $c^2=1+\beta^2$, the bulk metric reads
\bea
\label{Come2}
ds_6^2
&=&
\Big(1+\beta^2\big(1-\epsilon(z)^2\big)\Big)
dy^2
+(dz+ \beta s(y)\epsilon(z)
dy)^2
\\
&-&\Big(1+\big(\frac{\beta\epsilon(z)z}{\sqrt{1+\beta^2}}
+\sqrt{1+\beta^2}|y|\big)
\big(H+\frac{\dot{H}}{H}\big)
\Big)^2dt^2
+a^2\Big(1+\big(\frac{\beta\epsilon(z)z}{\sqrt{1+\beta^2}}
+\sqrt{1+\beta^2} |y|\big)
H\Big)^2\delta_{ij}dx^i dx^j.\nonumber
\eea
From the comparison with the ADM metric in Eq. (\ref{3}),
it is straightforward to read off
the lapse function, shift vector and 
induced metric components.
It is clear that the metric (\ref{Come2})
is invariant 
under the parity transformation
$y\to -y$ or $z\to -z$.
The assumption of the regularized profile of 
$\epsilon(z)$
is particularly important
to see the contributions from the 6D bulk
on the codimension-2 brane.
On the other hand,
it is enough to assume
the codimension-1 brane
as a distributional object, 
where
$\epsilon(z)$ and $\delta_{\epsilon}(z)$
are treated as the usual sign function 
and 
the delta function, respectively.

Clearly,
the bulk solution Eq. (\ref{Come2})
is 
included
in the metric class of Eq. (\ref{hom_6d_7})
with the replacements 
\bea
\label{hsp}
&&
N(t,y,z)^2=1+\beta^2\big(1-\epsilon(z)^2\big),\quad
u(t)=1,\quad
f_g(t,y,z)=1,\quad
f_n(t,y,z)=\beta,
\nonumber\\
&&
g_{tt}(t,y,z)
=-
\Big(1+\big(\frac{\beta\epsilon(z)z}{\sqrt{1+\beta^2}}
+\sqrt{1+\beta^2}|y|\big)
\big(H(t)+\frac{\dot{H}(t)}{H(t)}\big)
\Big)^2,
\nonumber\\
&&
g_{ij}(t,y,z)
=a(t)^2\Big(1+\big(\frac{\beta\epsilon(z)z}{\sqrt{1+\beta^2}}
+\sqrt{1+\beta^2} |y|\big)
H(t)\Big)^2\delta_{ij}.
\eea

\subsection{General cosmological equations on the codimension-2 brane}

We now derive the cosmological solutions
in the 4D spacetime.
Assuming metric (\ref{Come2}), the induced metric on the $y=0$ hypersurface is given by 
\bea
ds_5^2
=
dz^2
-\Big(1+\frac{\beta\epsilon(z)z}{\sqrt{1+\beta^2}}
\big(H+\frac{\dot{H}}{H}\big)
\Big)^2dt^2
+a^2
\Big(1+\frac{\beta\epsilon(z)z}{\sqrt{1+\beta^2}}
H\Big)^2
\delta_{ij}dx^i dx^j.
\label{come3}
\eea
Calculating the extrinsic and intrinsic curvature tensors (see Appendix 
A for details) and using the boundary Eq. (\ref{18}), we find the modified Friedmann equations on the codimension-2 brane,
\bea
\label{cosmic}
&&3M_4^2
H^2
=\rho
+\rho_{\rm eff},
\nonumber
\\
&&-M_4^2
\big(2\dot{H}+3H^2\big)
=p+p_{\rm eff},
\eea
where
$\rho$ and $p$ are
energy density and pressure of matter localized to
the codimension-2 brane,
and 
$\rho_{\rm eff}$ and $p_{\rm eff}$
represent
the effective energy density and pressure of
the dark component
\bea
\rho_{\rm eff}
&:=&\frac{6M_5^3\beta}{\sqrt{1+\beta^2}}
 H
-4M_6^4\int dz
\frac{\beta \delta_{\epsilon}(z)}
     {\sqrt{1+\beta^2(1-\epsilon(z)^2)}}
\Big(1+\frac{\beta\epsilon(z) z}{\sqrt{1+\beta^2}}
\big(\frac{\dot{H}}{H}+H\big)
\Big)^2,
\label{effcomp}
\nonumber\\
\\
p_{\rm eff}
&:=&
-\frac{6M_5^3\beta}{\sqrt{1+\beta^2}}
\Big(
H+\frac{\dot{H}}{3H}
\Big)
+4M_6^4
\int 
 dz
\frac{\beta\delta_{\epsilon}(z)}
     {\sqrt{1+\beta^2(1-\epsilon(z)^2)}}
\Big(1+\frac{\beta\epsilon(z) z}{\sqrt{1+\beta^2}}
H
\Big)^2.\nonumber
\eea
The effective equation of state is given by 
\bea
w_{\rm eff}&:=&\frac{p_{\rm eff}}{\rho_{\rm eff}}
\\
&=&
-1
-\frac{\dot{H}}{H}\frac{1}{\rho_{\rm eff}}
\Big[
\frac{2M_5^3\beta}{\sqrt{1+\beta^2}}
+4M_6^4\int dz
\frac{\beta\delta_{\epsilon}(z)}
     {\sqrt{1+\beta^2(1-\epsilon(z)^2)}}
\frac{\beta \epsilon(z)z}{\sqrt{1+\beta^2}}
\Big(
2
+\frac{\beta\epsilon(z) z}{\sqrt{1+\beta^2}}
\Big(
2H+\frac{\dot{H}}{H}
\Big)
\Big)
\Big].\nonumber
\label{eos}
\eea
Finally,
the Bianchi identity on the codimension-2 brane gives
the nonconservation law
\bea\dot{\rho}+3H\big(\rho+p\big)
&=&4 M_6^4
\int dz\frac{\beta \delta_{\epsilon}(z)}{\sqrt{1+\beta^2(1-\epsilon(z)^2)}}
\frac{\beta\epsilon(z) z}{\sqrt{1+\beta^2}} 
\Big\{
2
\frac{d}{dt}\Big(H+\frac{\dot{H}}{H}\Big)
\Big(
1
+\frac{\beta\epsilon(z) z}{\sqrt{1+\beta^2}} 
\Big(
H+\frac{\dot{H}}{H}
\Big)
\Big)
\nonumber\\
&+&3\dot{H}
\Big(
2
+\frac{\beta\epsilon(z) z}{\sqrt{1+\beta^2}} 
\big(
2H
+\frac{\dot{H}}{H}
\big)
\Big)
\Big\}.
\label{cons_general}
\eea
This may not be a surprising fact
since the inclusion of a codimension-2 brane thickness
is already 
somewhat beyond our theory \eqref{action}
where the codimension-2 brane is assumed to be an infinitesimally thin object.
In other words,
to see the nontrivial bulk effect on the brane dynamics,
it is necessary to include a finite brane thickness. 
For a de Sitter or Minkowski codimension-2 brane
where $H$ is constant,
the energy conservation law on the codimension-2 brane is 
satisfied even including a finite thickness.
From Eq. (\ref{cosmic}) with (\ref{effcomp}),
the Hubble parameter $H$ and its derivative $\dot{H}$
can be expressed in terms of the components of 
the energy-momentum tensor on the codimension-2 brane, $\rho$ and $p$.
On the other hand, 
combining the effective gravitational equation on the codimension-1 brane (\ref{11_1})
with (\ref{ext_5d}) and (\ref{int_5d}),
the time evolution of the energy-momentum tensor on the codimension-1 brane
$S_{ab}$
can be expressed only in terms of $H$ and $\dot{H}$,
and hence in terms of $\rho$ and $p$,
although the explicit form becomes quite complicated.
Therefore,
we do not have
the complete freedom to choose the energy-momentum tensor on the codimension-1 brane,
but instead 
it has to be tuned with that of the codimension-2 brane.


From now on, in the main text of this paper
we will focus on the cosmological dynamics on the codimension-2 brane.
We will discuss the dynamics on the codimension-1 brane, 
in more detail in Sec. 5,
where we will restrict the model parameters 
by imposing the null energy condition
as the criterion for the stability.

\subsection{Small brane thickness approximation}

In general it is impossible to
perform the integrals in Eq. (\ref{effcomp}) analytically.
Here we use an approximation in which the
codimension-2 brane thickness $\sigma$ is much smaller than 
the size of the cosmic horizon $H^{-1}$.
The details of the approximation are shown in Appendix 
B.
We then find
\bea
\label{eff_eqs}
\rho_{\rm eff}
&=&
\frac{6M_5^3\beta}{\sqrt{1+\beta^2}}
H 
-4M_6^4
\Big[\arctan (\beta)
+\sigma C(\beta) 
\Big(
H
+\frac{\dot{H}}{H}
\Big)
\Big],
\nonumber\\
p_{\rm eff}
&=&
-\frac{
6M_5^3\beta}{\sqrt{1+\beta^2}}
\Big(
H+\frac{\dot{H}}{3H}
\Big)
+4M_6^4
\Big[
\arctan (\beta)
+\sigma C(\beta)
 H
\Big],
\eea
where we have defined
\bea
\label{cbet}
C(\beta):=
-\frac{1}{\sqrt{1+\beta^2}}
+\frac{\sqrt{1+\beta^2}}{\beta}\arctan(\beta),
\eea
which
is a non-negative even function of $\beta$ that
vanishes at $\beta =0$
and monotonically increases towards $\frac{\pi}{2}$
as $|\beta|\to\infty$.
\begin{figure}
\begin{center}
\includegraphics[width=0.5\textwidth]{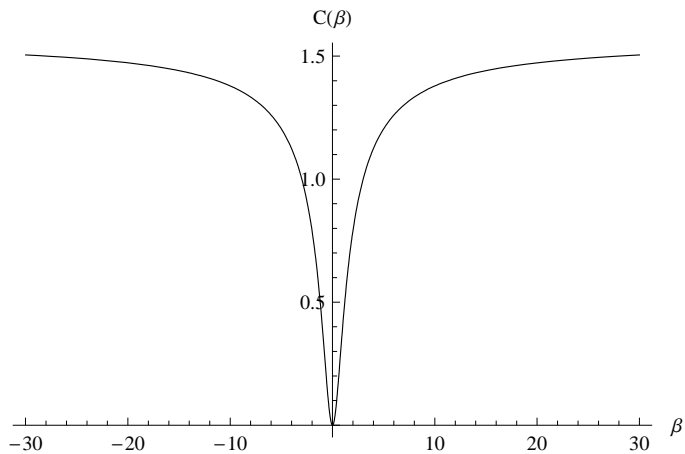}
\caption{
$C(\beta)$ as defined in Eq. (\ref{cbet}).}
\label{fig1}
\end{center}
\end{figure}
The conservation law on the codimension-2 brane Eq. (\ref{cons_general})
reduces to
\bea
\dot{\rho}+3H(\rho+p)
=4M_6^4 \sigma C(\beta)
\Big[\frac{d}{dt}\Big(\frac{\dot{H}}{H}\Big)
 +4\dot{H}
\Big].
\eea
Thus, due to a finite thickness
of the codimension-2 brane
there is always an energy exchange between
the bulk and the brane.
The effective equation of state (\ref{eos})
becomes
\bea
w_{\rm eff}
=
-1
-\frac{\dot{H}}{H^2}
\frac{\frac{2M_5^3\beta}{\sqrt{1+\beta^2}}
     +4M_6^4\sigma C(\beta)}
{\frac{6M_5^3\beta}{\sqrt{1+\beta^2}}
-\frac{4M_6^4}{H}
\big(\arctan(\beta)
+\sigma C(\beta)
\big(H+\frac{\dot{H}}{H}\big)
\big)}.
\eea
For $\big|\dot{H}\big|\ll H^2$
and since $\sigma H\ll 1$,
\bea
w_{\rm eff}
\simeq 
-1
-\frac{\dot{H}}{H^2}
\frac{\frac{M_5^3\beta}{\sqrt{1+\beta^2}}
     +2M_6^4\sigma C(\beta)}
{\frac{3M_5^3\beta}{\sqrt{1+\beta^2}}
-\frac{2M_6^4}{H}\arctan(\beta)}.
\eea


\section{De Sitter solutions and the features of degravitation}

In this section,
we 
give the solutions with Minkowski
and de Sitter codimension-2 branes
in the cascading model. 
In particular, 
we discuss
the possible features of degravitation.

\subsection{Minkowski codimension-2 brane}

The simplest solution is a Minkowski codimension-2 brane solution
which is realized by setting $H=0$ and $\dot{H}=0$
 in (\ref{Come2}).
In this case,
the 6D bulk metric is given by 
\bea
ds_6^2
&=&
\Big(1+\beta^2\big(1-\epsilon(z)^2\big)\Big)dy^2
+(dz+ \beta s(y)\epsilon(z)
dy)^2
-dt^2
+\delta_{ij}dx^i dx^j.
\label{Mink}
\eea
The codimension-2 brane is supported by
the tension $\rho=-p=\lambda$.
Modified Friedmann equations (\ref{cosmic})
reduce to the single equation
which relates the brane tension to the
bulk geometry,
\bea
\label{tune}
\lambda =4M_6^4\arctan (\beta),
\eea
where the left-hand side 
shows the deficit angle in the bulk.
For a Minkowski codimension-2 brane, $\tilde K_{ab}=0$,
and the codimension-1 brane geometry is also
the 5D Minkowski spacetime.
Since $|\arctan(\beta)|<1$,
the brane tension can be at most
of order $M_6^4$.
In Refs. \cite{gfree2,gfree},
it has been shown that the cascading gravity is 
ghost-free
if the codimension-2 brane tension satisfies the
following bound:
\bea
\lambda>\frac{2r_3}{3r_4}M_6^4.
\label{gf}
\eea
This condition is satisfied
as long as
two crossover scales satisfy
$r_3< r_4$.
Perturbations about the Minkowski codimension-2 brane solution (\ref{Mink})
and stability
have been explicitly analyzed in
the recent work Ref. \cite{gfree}.

\subsection{De Sitter codimension-2 brane and degravitation}

Now
we consider the de Sitter codimension-2 brane
solution where $a(t)=a_0 e^{H_0 t}$. 
In particular, we focus on the possible features of degravitation.
In this case 
the 6D bulk metric (\ref{Come2})
reduces to
\bea
ds_6^2
&=&
\Big(1+\beta^2\big(1-\epsilon(z)^2\big)\Big)dy^2
+(dz+ \beta s(y)\epsilon(z)
dy)^2
-\Big(1+\big(\frac{\beta\epsilon(z)z}{\sqrt{1+\beta^2}}
+\sqrt{1+\beta^2}
 |y|
\big)H_0\Big)^2dt^2
\nonumber\\
&+&a_0^2 e^{2H_0 t}
\Big(1+\big(\frac{\beta\epsilon(z)z}{\sqrt{1+\beta^2}}
+\sqrt{1+\beta^2} 
|y|
\big)H_0\Big)^2\delta_{ij}dx^i dx^j.
\label{dS}
\eea
Here
the codimension-2 brane can be supported only by 
the tension with $\rho=-p=\lambda$.
The effective cosmological equations (\ref{cosmic})
reduce to a single integral equation
\bea
\label{deSitter}
&&3M_4^2 H_0^2
=\lambda
+\rho_{\rm eff},
\eea
where
\bea
&&\rho_{\rm eff}
=-p_{\rm eff}
=
\frac{6M_5^3\beta }{\sqrt{1+\beta^2}}H_0
-4M_6^4 \int dz\frac{\beta\delta_{\epsilon} (z)}
        {\sqrt{1+\beta^2\big(1-\epsilon(z)^2\big)}}
\Big(1+\frac{\beta\epsilon(z)z}{\sqrt{1+\beta^2}}H_0\Big)^2.
\label{interm}
\eea

In the following, we obtain explicit de Sitter 3-brane solutions.
The solution to Eq. (\ref{deSitter})
gives the expansion rate of the de Sitter codimension-2 brane.
The solution of $H_0>0$ ($<0$) represents an expanding (contracting)
de Sitter universe in terms of the flat slicing.
In the case of $\lambda=0$, we discuss
some self-accelerating solutions.

\subsubsection{Recovering DGP solutions}

In the absence of the bulk gravity where $M_6\to 0$,
the modified cosmological equation on codimension-2 brane reads
\bea
3M_4^2 H_0^2
=\lambda
+\frac{6M_5^3\beta }{\sqrt{1+\beta^2}}H_0,
\eea
which corresponds to the cosmological equation in the DGP model with 
a codimension-2 brane tension.
The solution is then given by
\bea
\label{5d_sol}
H^{(\pm)}_0
=\frac{\frac{3M_5^3\beta}{\sqrt{1+\beta^2}}\pm 
\sqrt{\big(\frac{3M_5^3 \beta}{\sqrt{1+\beta^2}}\big)^2 
+3M_4^2\lambda}}
      {3M_4^2},
\eea
where $H_0^{(+)}> H_0^{(-)}$.
If $\lambda>0$,
irrespective of the sign of $\beta$
the solution of $H_0^{(+)}
$ 
represents the expanding de Sitter universe
and
that of $H_0^{(-)}
$
represents the contracting de Sitter universe.
If $-\frac{3M_5^6\beta^2}{(1+\beta^2)M_4^2}<\lambda<0$,
for $\beta>0$
both the solutions of $H_0^{(\pm)}
$ 
represent the expanding de Sitter universe
while
for $\beta<0$ 
both the solutions of $H_0^{(\pm)}
$ represent
the contracting de Sitter universe.
If $\lambda=0$,
for $\beta>0$
the solution of 
$H_0^{(+)}=\frac{2M_5^3\beta}{\sqrt{1+\beta^2}M_4^2}$
gives the self-accelerating solution of the DGP model,
while
for $\beta<0$
the solution of $H_0^{(+)}=0$ gives the normal branch
Minkowski brane solution of DGP.
But the self-accelerating solution in the DGP model
is known to be unstable.
The classification of solutions is shown in Table I.
\begin{table}
\begin{center}
{\scriptsize
\begin{tabular}{|c||c|c|c|
}
\hline
$\lambda$
&
$\beta$
&
$H_0^{(+)}$
&
$H_0^{(-)}$
\\
\hline
\hline
$\lambda>0$
&
$\beta>0$
&
Expanding
&
Contracting
\\
\cline{2-4}
&
$\beta<0$
&
Expanding
&
Contracting
\\
\hline
$\lambda=0$
 &
$\beta>0$
&
Expanding
&
Minkowski
\\
\cline{2-4}&
$\beta<0$
&
Minkowski
&
Contracting
\\
\hline
$-\frac{3M_5^6\beta^2}{M_4^2(1+\beta^2)}<\lambda<0$
 &
$\beta>0$
&
Expanding
&
Expanding
\\
\cline{2-4}&
$\beta<0$
&
Contracting
&
Contracting
\\
\hline
\end{tabular}
}
\label{table_1}
\end{center}
\caption{\baselineskip 14pt
In this table the classification of solutions
 is shown.
The terms ``expanding'' and ``contracting''
denote the expanding and contracting de Sitter universe,
respectively.
The term ``Minkowski'' denotes
the Minkowski codimension-2 brane.
Note that the same terms are used in the subsequent tables.
}
\end{table}

\subsubsection{Degravitation features
in the limit of zero brane thickness}

In this section we study the effective cosmological equations (\ref{deSitter}) in the limit of zero thickness of the codimension-2 brane. In this case Eq. (\ref{deSitter}) becomes
\bea
\label{H0}
3M_4^2 H_0^2
=\tilde\lambda
+\frac{6M_5^3\beta }{\sqrt{1+\beta^2}}H_0,
\quad
\tilde\lambda:=
\lambda-4M_6^4\arctan (\beta).
\eea
The solution is simply given by 
Eq. (\ref{5d_sol}) classified in Table 1,
with replacement of $\lambda$
with $\tilde\lambda$. In the case of  $\lambda=0$,
for $\beta>0$
both the solutions of $H_0^{(\pm)}$
give rise to a self-accelerating universe 
for $M_5^6 
>\frac{4M_4^2M_6^4(1+\beta^2)\arctan \beta}{3\beta^2}$.
For $\beta<0$
only the solution of $H_0^{(+)}
$ provides the self-accelerating universe,
while $H_0^{(-)}
$ leads to a contracting universe.
However,
these self-accelerating solutions
might be unstable against perturbations
due to the possible existence of a ghost mode
because they do not satisfy the condition (\ref{gf}).

Now we argue the possible connections of our solutions
to the degravitation expected in the cascading model.
We focus on the case
of $\tilde \lambda <0$
and $\beta>0$
where both solutions
represent the expanding de Sitter universe.
In the limiting case of 
$3M_5^6\beta^2 \gg
-
(1+\beta^2)M_4^2{\tilde\lambda}$, we have
\bea
\frac{H_0^{(-)}}{H_0^{(+)}}
\simeq 
-\frac{(1+\beta^2)}{12\beta^2 }\frac{r_3}{r_4}
\frac{\tilde \lambda}{M_6^4}
\ll 1,\label{ratio}
\eea
for $\frac{\tilde \lambda}{M_6^4}=O(1)$
as long as $r_3\ll r_4$.
The physical meaning of Eq. (\ref{ratio})
is clearly understood as follows:
The $H_0^{(-)}$ solution with the tension
$\frac{2}{3}\frac{M_4^2M_6^8}{M_5^6}
<\lambda
<4M_6^4\arctan\beta$,
which is expected to be ghost-free,
gives a much smaller expansion rate
than one in the self-accelerating branch
in the 5D DGP model
with an expansion rate of order $H_0^{(+)}$.
In addition,
rewriting the $H_0^{(-)}$ solution in the same limit
\bea
\label{hm}
H_0^{(-)}{}^2
\simeq
\frac{1}{3M_4^2}
\Big(\frac{r_3}{r_4}\Big)
|\tilde\lambda|
\ll 
\frac{1}{3M_4^2}
|\tilde\lambda|,
\eea
shows that
the effect of
the effective vacuum energy
$\tilde\lambda$
on the Hubble expansion rate
is suppressed
by the ratio $\frac{r_3}{r_4}$
in comparison with 
the naive expectation
from the ordinary 4D cosmology. This implies a deep connection of
our results with the degravitation idea
and shows that cascading model can provide a mechanism that could
support a small expansion rate 
in the presence of a large
cosmological constant i.e., tension.  
Moreover,
although the suppression in the 6D model
is not enough to explain the fine-tuning problem 
for
obtaining a tiny expansion rate which is suggested from observations,
the cascading gravity model may be extendable to higher-dimensional
cases,
which may lead to more crossover scales
$r_i:=\frac{M_{i+1}^{i-1}}{M_{i+2}^{i}}$ ($i=5,6,7,\cdots$)
with $ r_{i}\ll r_{i+1}$.
Then we can expect that the fine-tuning problem
will be alleviated more than in the 6D model.

\subsubsection{Degravitation features with a small brane thickness}

In general, 
it is impossible to solve the integral 
Eq. (\ref{deSitter}) with (\ref{interm}) analytically
when a finite thickness of the
codimension-2 brane is taken into consideration.
Here, we use the small thickness approximations
given in the previous section and in Appendix B.
Using (\ref{eff_eqs}),
Eq. (\ref{deSitter}) reduces to the 
quadratic equation for $H_0$
which leads to the solutions
\bea
 H_0
=H_{0}^{(\pm)}
&:=&\frac{1}{3\sqrt{1+\beta^2}M_4^2}
\Big[
3M_5^3\beta
-2\sqrt{1+\beta^2} \sigma C(\beta) M_6^4
\nonumber\\
&\pm&
\sqrt{
\big(3M_5^3\beta
-2\sqrt{1+\beta^2} \sigma C(\beta) M_6^4\big)^2
+
3M_4^2 
(1+\beta^2)
\tilde\lambda
}
\Big].
\eea
There is no physical solution for
\bea
\label{real}
\tilde \lambda<
{\tilde\lambda}_{\ast}
:=
-\frac{(3M_5^3\beta-2\sqrt{1+\beta^2}\sigma C(\beta)M_6^4)^2}
      {3(1+\beta^2)M_4^2}.
\eea
If  $\tilde\lambda>0$
the solution of $H_0^{(+)}$
always represents the expanding de Sitter universe.
If $
{\tilde\lambda}_{\ast}
<\tilde\lambda<0$ and $\beta>\frac{2\sqrt{1+\beta^2}\sigma C(\beta) M_6^4}{3M_5^3}$
both solutions of $H_0^{(\pm)}$ represent
the expanding de Sitter universe.
In the absence of the brane tension, $\lambda=0$, 
we can obtain self-accelerating solutions. 
If $\beta>0$,
for $3M_5^3\beta>2\sqrt{1+\beta^2} \sigma C(\beta) M_6^4$,
both solutions of $H_0^{(\pm)}$ give 
self-accelerating universes, while
if $\beta<0$ only the solution of $H_0^{(+)}$
gives a self-accelerating universe.
However,
these self-accelerating solutions
might be unstable against perturbations
since they do not satisfy the condition (\ref{gf}).
The classification of solutions is shown in Table II.

Finally, we argue the connections
of the $H_0^{(-)}$ solution
with the idea of 
degravitation.
For $\beta>0$
and $3M_5^3\beta
>2\sqrt{1+\beta^2} \sigma C(\beta) M_6^4$,
in the limit of 
 $
\tilde\lambda\gg 
\tilde \lambda_{\ast}$,
we obtain
\bea
\label{hm3}
\frac{H_0^{(-)}}{H_0^{(+)}}
\simeq
-
\frac{1}{
\big(1
-\frac{2\sqrt{1+\beta^2}\sigma C(\beta)}
      {3\beta r_4}
\big)^2}
\frac{1+\beta^2}{12\beta^2}
\Big(\frac{r_3}{r_4}\Big)
\frac{\tilde\lambda}{M_6^4}
\ll  1,
\eea
and 
\bea
\label{hm2}
H_0^{(-)}{}^2
\simeq
\frac{1}{3M_4^2}
\Big(\frac{r_3}{r_4}\Big)
\frac{1}{
\big(1
-\frac{2\sqrt{1+\beta^2}\sigma C(\beta)}
      {3\beta r_4}
\big)^2}
|\tilde\lambda|,
\eea
for $\frac{{\tilde \lambda}}{M_6^4}=O(1)$
and $\frac{r_3}{r_4}\ll 1$.
These equations differ from Eq. (\ref{hm}) by the factor $\big(1
-\frac{2\sqrt{1+\beta^2}\sigma C(\beta)}
      {3\beta r_4}
\big)^{-2}$which depends on the thickness of the brane. Since  $r_4\gg \sigma$,
where $r_4$ could be macroscopic and 
$\sigma$ is microscopic, the effects of the brane thickness
are small. Therefore, one could still conclude that 
 the Hubble expansion rate is much smaller than the one in
 ordinary 4D cosmology, namely,
$H_0^{(-)}{}^2\ll \frac{1}{3M_4^2}|\tilde\lambda|$.
Thus, 
irrespective of the inclusion of a small thickness, the cascading model exhibits features of degravitation.

\begin{table}

\begin{center}
{\scriptsize
\begin{tabular}{|c||c|c|c|
}
\hline
$\lambda$
&
$\beta$
&
$H_0^{(+)}$
&
$H_0^{(-)}$
\\
\hline
\hline
$\tilde \lambda>0$
&
$\beta>
\frac{2\sqrt{1+\beta^2}\sigma C M_6^4}{3M_5^3}
$
&
Expanding
&
Contracting
\\
\cline{2-4}
&
$\beta<
\frac{2\sqrt{1+\beta^2}\sigma C M_6^4}{3M_5^3}
$
&
Expanding
&
Contracting
\\
\hline
$\tilde \lambda=0$
 &
$\beta>\frac{2\sqrt{1+\beta^2}\sigma C M_6^4}{3M_5^3}
$
&
Expanding
&
Minkowski
\\
\cline{2-4}&
$\beta<\frac{2\sqrt{1+\beta^2}\sigma C M_6^4}{3M_5^3}
$
&
Minkowski
&
Contracting
\\
\hline
$
{\tilde \lambda}_{\ast}
<\tilde\lambda<0$
 &
$\beta>
\frac{2\sqrt{1+\beta^2}\sigma C M_6^4}{3M_5^3}
$
&
Expanding
&
Expanding
\\
\cline{2-4}&
$\beta<
\frac{2\sqrt{1+\beta^2}\sigma C M_6^4}{3M_5^3}
$
&
Contracting
&
Contracting
\\
\hline
\end{tabular}
}
\label{table_1}
\end{center}
\caption{\baselineskip 14pt
Classification of solutions
with a codimension-2 brane thickness.
}
\end{table}

In the next section, we will discuss
the properties of the energy-momentum tensor of  matter field
localized on the codimension-1 brane.
Our criterion for the stability is
whether the solution satisfies the null energy condition.
We will also investigate 
whether
it is 
compatible with the expected ghost-free condition 
discussed in this section.

\section{Dynamics on the codimension-1 brane and stability}

\subsection{Components of the energy-momentum 
tensor on the codimension-1 brane}

Here we shall briefly mention the dynamics on the codimension-1 brane.
The nonvanishing components of
the extrinsic curvature tensors
on the codimension-1 brane
are given in Appendix A.
The components of the energy-momentum tensor
on the codimension-1 brane $S_{ab}$
are determined by 
the gravitational equations (\ref{9}).
From Eq. (\ref{9}) and the fact that ${\tilde K}_{z\mu}=0$
along the codimension-1 brane ($\mu=t,i$), we get
$S_{z\mu}$=0.
Thus, it is straightforward to check that 
the components of the extrinsic curvature tensor satisfy
the condition $\nabla_b({\tilde K}^b_a-\delta^b_a 
{\tilde K})=0$.
Hence,
the conservation law equation $\nabla_bS^b{}_a=0$
is also satisfied.
Similarly in the case of a codimension-2 brane
without a thickness,
since all the components of the 6D Einstein tensor
vanish everywhere in the bulk,
there is no energy exchange between the
codimension-2 brane and the bulk.
But the inclusion of a finite thickness
leads to an energy exchange between them
if the codimension-2 brane geometry
is not Minkowski or de Sitter symmetry.

For a de Sitter codimension-2 brane, 
using Eq. (\ref{9})
together with  (\ref{dS_ext}) and (\ref{dS_int}),
the energy-momentum tensor of matter localized to the 
codimension-1 brane is given by 
\bea
S^{\mu}{}_{\nu}
&=&\frac{
     -3H_0^2 M_5^3
     +6M_6^4 H_0 \big(\sqrt{1+\beta^2}+\beta H_0 z\big)}     
  {\big(\sqrt{1+\beta^2}+\beta H_0 z\big)^2}
\delta^{\mu}{}_{\nu}
=M_6^4 H_0
\frac{-3H_0 r_4+6(\sqrt{1+\beta^2}+\beta H_0 z)}
     {\big(\sqrt{1+\beta^2}+\beta H_0 z\big)^2}
\delta^{\mu}{}_{\nu},
\nonumber\\
S^{z}{}_{z}
&=&\frac{
     -6H_0^2 M_5^3
     +8M_6^4 H_0 \big(\sqrt{1+\beta^2}+\beta H_0 z\big)}     
  {\big(\sqrt{1+\beta^2}+\beta H_0 z\big)^2}
=M_6^4 H_0
\frac{-6H_0 r_4+8(\sqrt{1+\beta^2}+\beta H_0 z)}
     {\big(\sqrt{1+\beta^2}+\beta H_0 z\big)^2},
\label{smunu}
\eea 
The five-dimensional energy density and pressures
are defined by 
$\rho_5=-S^t{}_t$,
$p_5=\frac{1}{3}S^i{}_i$
and 
$p_{5,z}=S^z{}_z$.
The codimension-1 brane can be supported by 
matter which has anisotropic pressure
$p_5\neq p_{5,z}$.
Note that they are
regular in the codimension-2 brane limit of $z\to 0$.
In this section,
we focus on the expanding 
four-dimensional de Sitter universe.

As the criterion for the stability of the codimension-1 brane,
we will impose the null energy condition
for the energy momentum tensor, $S_{ab}n^an^b\geq 0$,
where $n^a$ is an arbitrary null vector field on the codimension-1 brane.
In our case, the null energy condition is given by 
\bea
\rho_5+p_5\geq 0,\quad
\rho_5+p_{5,z}\geq 0.
\label{nec}
\eea 
In the rest of this paper,
we will focus on the behavior of
the energy-momentum tensor of matter 
localized on the codimension-1 brane
in the various limiting cases.
Imposing the null energy condition,
we will discuss the restrictions on the model parameters.

\subsection{Behaviors of the energy-momentum tensor}

In this subsection, 
we discuss the behavior of the energy-momentum tensor
and restrict the model parameters.
First,
around the codimension-2 brane $z\to 0+$,
the five-dimensional energy density and pressures
satisfy
\bea
\rho_5+p_5=0, \quad
\rho_5+p_{5,z}
=M_6^4 H_0
 \frac{-3r_4 H_0+2\sqrt{1+\beta^2}}
      {1+\beta^2}.
\eea
Imposing the null energy condition Eq. (\ref{nec}),
for $H_0>0$ we obtain an upper bound on $H_0$:
\bea
H_0\leq \frac{2\sqrt{1+\beta^2}}{3r_4}.
\label{nec_Hubble}
\eea

Second, 
we investigate the behavior
away from the codimension-2 brane. 
For $\beta<0$, 
the energy-momentum tensor diverges 
at $z=-\frac{\sqrt{1+\beta^2}}{H_0 \beta}$,
where a singularity on the codimension-1 brane 
appears.
Thus we have to impose $\beta>0$,
for which it is regular on the codimension-1 brane ($z>0$).
In the limit $z\to \infty$,
\bea
\rho_5+p_5 \big|_{z\to\infty} =0,\quad
\rho_5+p_{5,z} \big|_{z\to\infty} 
\to\frac{2M_6^4}{\beta z}>0.
\eea
Thus for $\beta>0$,
and if Eq. (\ref{nec_Hubble}) is satisfied,
the null energy condition is satisfied
over the codimension-1 brane.

\subsection{
Restrictions on the parameters}

Finally, 
we apply the above constraints
to the solutions obtained in the zero thickness limit
in Sec. 4.2.2.
As we have seen in Sec. 4.2.3,
as long as the thickness of the 
codimension-2 brane 
is negligibly small,
corrections due to a finite thickness
are also negligible.
Thus 
it is enough to focus on this limit. 
The solutions to Eq. (\ref{H0})
are given by 
\bea
H_{0}^{(\pm)}
=
\frac{1}{3r_3}
\Big[
\frac{3\beta}{\sqrt{1+\beta^2}}
\pm
\sqrt{
\Big(
\frac{3\beta}{\sqrt{1+\beta^2}}
\Big)^2
+\frac{3{\tilde\lambda}}
      { M_6^4}
 \frac{r_3}{r_4}
}
\Big].
\eea
We impose $\beta>0$,
for which both solutions of $H_{0}^{(\pm)}$
describe the expanding de Sitter universes.
In order to make both solutions
real and positive,
we have to impose a condition on the tension
\bea
0
<
\frac{\big(-{\tilde\lambda}\big)}{M_6^4}
<
\frac{3\beta^2}
     {1+\beta^2}
\frac{r_4}{r_3}.
\label{tens1}
\eea

If $|\tilde\lambda|\ll 
\frac{3\beta^2}{1+\beta^2}
 \frac{r_4}{r_3}M_6^4$
with ${\tilde\lambda}=O(M_6^4)$,
the solutions $H_0^{(\pm)}$ approach
\bea
H_0^{(+)}
\simeq 
\frac{2\beta}{\sqrt{1+\beta^2}}
\frac{1}{r_3},
\quad
H_0^{(-)}
\simeq 
\frac{\sqrt{1+\beta^2}}{6\beta}
\frac{|{\tilde\lambda}|}{M_6^4}
\frac{1}{r_4},
\eea
respectively.
Concerning the solutions
of $H_0^{(+)}>\frac{2\beta}{\sqrt{1+\beta^2}}\frac{1}{r_3}$,
if $r_3\ll r_4$,
they cannot satisfy Eq. (\ref{nec_Hubble}).
This condition becomes more severe
than that of the expected ghost-free condition
in the previous section
that suggests
the stability of the solution of $H_0^{(+)}$ 
for $\beta>0$.

Concerning the solutions of $H_0^{(-)}$, 
because of $H_0^{(-)}> \frac{\sqrt{1+\beta^2}}{6\beta}
                  \frac{|\tilde{\lambda}|}{M_5^3}$
and Eq. (\ref{nec_Hubble}),
a solution which satisfies the null energy condition 
can be obtained
for the codimension-2 brane tension
$\frac{(-{\tilde\lambda})}{M_6^4}<4\beta$,
where we have used $r_3\gg r_4$.
Combined with Eq. (\ref{tens1}),
a tighter bound on the brane tension is obtained
\bea
\label{tens_bound_sev}
0<
\frac{(-{\tilde\lambda})}{M_6^4}
<4\beta.
\eea
For the brane tension of (\ref{tens_bound_sev}),
the solution of $H_0^{(-)}$ would be stable,
and $H_0^{(-)}\ll H_0^{(+)}$.
On the other hand,
as we mentioned in Sec. 4.2.2,
for $(-{\tilde\lambda})>0$ and $\beta>0$,
the expected ghost-free condition 
$\frac{2}{3}\frac{r_3}{r_4}<\lambda<4M_6^4\arctan \beta$
is rewritten as
\bea
\label{minkowski_bound_sev}
0<
\frac{(-\tilde\lambda)}{M_6^4}
<4\arctan(\beta)-\frac{2r_3}{3r_4}
\simeq 
4\arctan(\beta)
\simeq 
4\beta
\eea
for $0<\frac{r_3}{r_4}\ll \beta<1$.
Thus,
for the branch of $H_0^{(-)}$,
the condition which we have obtained (\ref{tens_bound_sev})
is approximately consistent with 
the expected ghost-free condition (\ref{minkowski_bound_sev}).


\section{Conclusion}

We have presented a
formulation
of the nonlinear effective gravitational theory
in the 6D cascading gravity model.
This model is a higher-dimensional extension
of the 5D DGP braneworld model.
In the simplest 6D model
we are living on 
a codimension-2 brane
that is located on a
codimension-1 brane embedded into a 6D bulk.
The particularly interesting expectations
are that 
this model
may exhibit a degravitation
where the gravitational force falls off
sufficiently fast,
and 
is also free from a ghost instability 
if the tension of the codimension-2 brane satisfies a bound.
An important aim for presenting our formulation is to 
see whether in reality
the idea of degravitation
could work for the cascading model,
through an explicit investigation of
cosmological solutions.
The gravitational equations on the codimension-2 brane 
are composed of the contributions
of matter on the codimension-2 brane,
the induced gravity on the codimension-1 brane
and the gravity in the 6D bulk.
The bulk contribution is given
by integrating over the sixth direction across the codimension-2 brane
along the codimension-1 brane.
After the derivation of the general equations of motion in the cascading model,
we applied them to cosmology
where the codimension-2 brane geometry is
described by a flat Friedmann-Robertson-Walker universe,
and we obtained the modified Friedmann equations.
In the zero thickness limit of the codimension-2 brane,
the bulk contribution becomes an effective cosmological constant
whose sign depends on the sign of $\beta$.
A finite thickness, however, 
leads to an energy exchange between the bulk and the codimension-2 brane,
except for the cases
where the codimension-2 brane geometry is exactly Minkowski or de Sitter.
On the other hand,
there is no energy exchange
from or into the codimension-1 brane in any case.
Note that 
our solutions 
are obtained 
in a different setup
from that in the original
cascading gravity model \cite{6,gfree2},
in the sense that 
the codimension-1 brane
contains  
matter fields other than the pure tension.
It turns out that 
in our model
the energy-momentum tensor on the codimension-1 brane
is not the one which
was motivated from a simple field theory model,
like a scalar field.
It is also interesting to compare cosmological behaviors in
the cascading gravity model with those in
the six-dimensional intersecting brane model 
with the induced gravity terms \cite{kalo,ckt,ckt2}.
In this model,
the second codimension-1 brane
intersects the first codimension-1 brane
at the position of the codimension-2 brane.
For a de Sitter codimension-2 brane with $\dot{H}=0$,
the cascading gravity model exhibits similar behaviors to the intersecting brane model,
in the sense that in the thin codimension-2 brane limit
the bulk contribution induces an 
effective cosmological constant on the codimension-2 brane 
\cite{ckt}.
For a more general 
homogeneous and isotropic codimension-2 brane with $\dot{H}\neq 0$,
some differences arise:
in the intersecting brane model
the energy-momentum of matter on the codimension-2 brane
is not conserved except for the case where the codimension-1 branes are at a right angle
\cite{ckt2},
while in the cascasding gravity model
it is conserved
if we ignore the thickness of the codimension-2 brane.

Finally, we have discussed the 
Minkowski and de Sitter codimension-2 brane solutions.
The Minkowski codimension-2 brane solution is realized 
if both the bulk and codimension-1 brane are empty
and the codimension-2 brane tension takes 
a particular value determined by the bulk geometry.
In the de Sitter brane solutions,
the bulk gravity effect gives rise to a new 
branch of the solution
which can give an expanding de Sitter codimension-2 brane solution
that is expected to be stable
and has a much smaller expansion rate
than in the original DGP model;
this leads to
alleviation of the fine-tuning problem.
We have also shown some de Sitter solutions with the 
small expansion rate which could have deep connections
with the idea of degravitation. 
This result shows that the cascading model provides a dynamical mechanism to resolve the cosmological constant problem.
We have also argued the stability of our model,
in terms of whether the energy-momentum tensor
on the codimension-1 brane
satisfies the null energy condition.
It restricts the model parameters severely.
Among two branches of the de Sitter solutions,
the branch which has the smooth limit to the self-accelerating 
solution in the 5D DGP model 
cannot satisfy the null energy condition.
On the other hand,
the above new branch solution 
can satisfy it,
if the tension satisfies (\ref{tens_bound_sev}).
This bound from the null energy condition 
is approximately consistent 
with the expected ghost-free condition in \cite{7,gfree2}.

Before closing this paper,
we would like to comment more on connections  with the degravitation \cite{6,degra}.
As shown in the effective Friedmann equation Eq. (\ref{hm}),
the contribution of 
the effective cosmological constant
$\tilde\lambda$
to the Hubble expansion rate
is suppressed
by the ratio $\frac{r_3}{r_4}$
in comparison with 
the naive expectation from the ordinary 4D cosmology.
This result shows that cascading gravity model can provide a mechanism that could
support a small expansion rate 
in the presence of a large effective cosmological constant.  
The degravitation \cite{6,degra}, however,
requires that the cosmological
constant is completely decoupled from gravity.
In such a sense
our model is not sufficient to provide the complete degravitation.
But, as argued below \eqref{hm},
the cascading gravity model may be extendable to higher-dimensional cases,
which leads to more crossover scales
$r_i:=\frac{M_{i+1}^{i-1}}{M_{i+2}^{i}}$ ($i=5,6,7,\cdots$) with $ r_{i}\ll r_{i+1}$.
Thus we expect that more suppression factors provided by the combinations of the new crossover scales
will appear
in the right-hand side of the effective Friedmann equation
as the number of the total spacetime dimension increases,
and 
the effect of the cosmological constant on the Hubble expansion rate
would be weakened drastically.
The other important feature of the degravitation 
is the scale dependence of the effective gravitational coupling.
At the linearized level, 
results supporting the scale dependence have been obtained in \cite{7,gfree2}.
In the case of the homogeneous and isotropic cosmology,
the assumptions in deriving \eqref{hm}
were $3M_5^6\beta^2 \gg
(1+\beta^2)M_4^2 |{\tilde\lambda}|$
and 
$\frac{|\tilde \lambda|}{M_6^4}=O(1)$
with $r_3\ll r_4$.
In the early universe
when the cosmic energy density is much larger than $|\tilde\lambda|$,
the correction terms from the codimension-1 brane and the 6D bulk would be negligible.
Thus we expect that the evolution of the early universe, e.g., inflation,
will not be affected by them.

\section*{Acknowledgement}
P.M. would like to thank Claudia de Rham, Cristiano Germani, Kurt Hinterbichler, Stefan Hofmann, Justin Khoury, 
and Florian Niedermann for helpful discussions. 
P.M. would also like to thank Olaf Hohm for useful comments on the first draft of the paper and the University of Geneva, where some part of this work was done, for hospitality. 
The work of P.M. was supported by the Alexander von Humboldt Foundation. 
The work of M.M. was supported 
by Grant-in-Aid for Young Scientists (B) of JSPS Research,
under Contract No. 24740162,
and by the FCT-Portugal through the Grant No. SFRH/BPD/88299/2012.

\appendix

\section{Components of tensors}

\subsection{For a general cosmological brane}

On the codimension-1 brane,
using metric (\ref{Come2}) with the identification Eq. (\ref{hsp})
and following the definition of Eq. (\ref{11}),
we obtain
the components
of the extrinsic curvature ${\tilde K}_{ab}$
except for the contribution of the codimension-2 brane:
\bea
&&{\tilde K}^t{}_t
=
\frac{1}{N}
\frac{
s(y)
\big(H+\frac{\dot{H}}{H}\big)}
{\sqrt{1+\beta^2}
+\big(\beta\epsilon(z)z+(1+\beta^2)
|y|\big)
\big(H+\frac{\dot{H}}{H}\big)}
\nonumber\\
&&{\tilde K}^i{}_j
=
\frac{\delta^i{}_j}{N}
\frac{
s(y) H}
{\sqrt{1+\beta^2}
+\big(\beta\epsilon(z)z+(1+\beta^2)
|y|\big)H},
\nonumber\\
&&{\tilde K}_{zz}
= {\tilde K}_{zt}={\tilde K}_{zi}=0.
\label{ext_c1}
\eea
In the limit of $y \to 0+$,
noting that  $N=1$ off the codimension-1 brane $y\neq 0$,
we obtain
the nonvanishing components of the 
extrinsic curvature tensor 
except for the contribution of the codimension-2 brane
\bea
{\tilde K}^t{}_t\to 
\frac{
H+\frac{\dot{H}}{H}}
     {\sqrt{1+\beta^2}
+\beta \epsilon (z) z 
\big(H+\frac{\dot{H}}{H}\big)},
\quad
{\tilde K}^i{}_j\to 
\frac{H}
     {\sqrt{1+\beta^2}+\beta \epsilon(z) z H}\delta^i{}_j,
\eea
and hence the nonvanishing components of 
the combination appearing in the junction condition
\bea
\label{ext_5d}
{\tilde K}^t{}_t-{\tilde K}
&\to&
-\frac{3H}{\sqrt{1+\beta^2}+\beta\epsilon(z) z H},
\quad
{\tilde K}^i{}_j-\delta^i{}_j 
{\tilde K}
\to 
-\Big(
\frac{
H+\frac{\dot{H}}{H}}
{\sqrt{1+\beta^2}+\beta \epsilon (z)z
\big(H+\frac{\dot{H}}{H}\big)}
+\frac{2H}{\sqrt{1+\beta^2}+\beta\epsilon(z) z H}
\Big)
\delta^i{}_j,
\nonumber\\
{\tilde K}^z{}_z- {\tilde K}
&\to &
-\Big(
\frac{H+\frac{\dot{H}}{H}}
{\sqrt{1+\beta^2}
+\beta\epsilon(z) z
\big(H+\frac{\dot{H}}{H}\big)}
+\frac{3H}
{\sqrt{1+\beta^2}+\beta\epsilon(z) zH}\Big).
\eea

Given the 5D metric (\ref{come3}),
following the definition of \eqref{14a} with 
\bea
&&{\cal N}=1,\quad {\cal N}^t={\cal N}^i =0,
\quad
g_{tt}
=-\Big(1+\frac{\beta\epsilon(z)z}{\sqrt{1+\beta^2}}
\big(H+\frac{\dot{H}}{H}\big)
\Big)^2,
\quad
g_{ij}
=
a(t)^2
\Big(1+\frac{\beta\epsilon(z)z}{\sqrt{1+\beta^2}}
H\Big)^2
\delta_{ij},
\eea
the nonvanishing components of
${\cal K}^{\mu}{}_{\nu}$ defined
on the codimension-1 brane ($y=0$) are given by
\bea
\mathcal{K}^{t}{}_t
&=&
\Big(1
+\frac{\beta\epsilon(z)z}{\sqrt{1+\beta^2}}
\big(H+\frac{\dot{H}}{H}\big)
\Big)^{-1}
\frac{\beta \big(\epsilon(z)+2z\delta_\epsilon(z)\big) 
}{\sqrt{1+\beta^2}}
\big(H+\frac{\dot{H}}{H}\big),
\nonumber\\
\mathcal{K}^{i}{}_j
&=&
\delta^i{}_j
\Big(1
+\frac{\beta\epsilon(z)z}{\sqrt{1+\beta^2}}
H\Big)^{-1}
\frac{\beta \big(\epsilon(z)+2z\delta_\epsilon(z)\big)}
{\sqrt{1+\beta^2}}
 H,\quad
\mathcal{K}_{ti}=0.
\eea
Since in the induced metric \eqref{come3}
$\epsilon(z)$ dependence appears in the form of $\epsilon (z) z$,
the diagonal components of 
${\cal K}^{\mu}{}_{\nu}$
are proportional to $2\delta_\epsilon (z) z+ \epsilon(z)$.
But on the codimension-1 brane,
the codimension-2 brane is a codimension-1 object and 
the thin brane limit [$\epsilon(z)\to s(z)$ and $\delta_\epsilon (z)\to \delta(z)$]
can be smoothly taken.
Noting $z\delta(z)\to 0$, we obtain
\bea
\mathcal{K}^{t}{}_t
&\to&
\Big(1
+\frac{\beta\epsilon(z)z}{\sqrt{1+\beta^2}}
\big(H+\frac{\dot{H}}{H}\big)
\Big)^{-1}
\frac{\beta}{\sqrt{1+\beta^2}}
\big(H+\frac{\dot{H}}{H}\big)
s(z),
\nonumber\\
\mathcal{K}^{i}{}_j
&\to&
\delta^i{}_j
\Big(1
+\frac{\beta\epsilon(z)z}{\sqrt{1+\beta^2}}
H\Big)^{-1}
\frac{\beta}
{\sqrt{1+\beta^2}}
 H s(z),\quad
\mathcal{K}_{ti}=0.
\eea
Thus, there is no explicit dependence on $\delta (z)$.
In the codimension-2 brane limit of $z\to 0+$
\bea
\mathcal{K}^t{}_t
&\to&
\frac{\beta }{\sqrt{1+\beta^2}}
\big(H+\frac{\dot{H}}{H}\big),
\quad
\mathcal{K}^{i}{}_j
\to 
\delta^i{}_j
\frac{\beta}{\sqrt{1+\beta^2}}H.
\label{ext_3brane}
\eea

The components of the five-dimensional 
Einstein tensor on the codimension-1 brane are given by
\bea
{}^{(5)}
G^z{}_z
&=&\frac{3(A^2-1)H}
       {\big(1+H Az\big)^2(H+ (H^2+\dot{H})Az)}
\Big(
2H(H^2+\dot{H})Az
+2H^2
+\dot{H}
\Big),
\nonumber\\
{}^{(5)}G^t{}_t
&=&
\frac{3(A^2-1)H^2}
     {\big(1+H Az\big)^2},
\nonumber\\
{}^{(5)}G^i{}_j
&=&\frac{H (A^2-1)}
{\big(1+H Az\big)^2
(H+(H^2+\dot{H})Az)}
\Big(
 3H(H^2+\dot{H}) Az
+2\dot{H}
+3H^2
\Big)\delta^i{}_j,
\label{int_5d}
\eea
where we have defined  $A:=\frac{\beta}{\sqrt{1+\beta^2}}$.

%

On the codimension-2 brane,
the nonvanishing components of the Einstein tensor  are given by 
\bea
{}^{(4)}G^t{}_t=
-3H^2
,\quad
{}^{(4)}G^{i}{}_j
= -\delta^i{}_j
\big(
3H^2+2\dot{H}
\big).
\label{4d_Eins_cosmo}
\eea
The homogeneity and isotropy of the geometry on the
codimension-2 brane
restricts the form of 
the energy-momentum tensor
to have only the diagonal components
\bea
\label{3_matter}
T^t{}_t=
-\rho,\quad
T^{i}{}_j
 =p 
  \delta^{i}{}_{j}.
\eea

\subsection{For a de Sitter codimension-2 brane}

For a de Sitter codimension-2 brane
Eqs. (\ref{ext_5d}) reduce to
\bea
{\tilde K}^\mu{}_{\nu}
-\delta^{\mu}{}_{\nu}{\tilde K}
=
-\frac{3H_0}
{\sqrt{1+\beta^2}+\beta z H_0}\delta^{\mu}{}_{\nu},
\quad
{\tilde K}^z{}_z-{\tilde K}
=
-\frac{4H_0}
{\sqrt{1+\beta^2}+\beta z H_0},
\label{dS_ext}
\eea
where $H_0$ is given by the solution of Eq. (\ref{H0}).
On the other hand, from Eq. (\ref{int_5d})
\bea
{}^{(5)}G^{\mu}{}_{\nu}
&=&-\frac{3H_0^2}
       {\big(\sqrt{1+\beta^2}+\beta H_0 z\big)^2}
\delta^{\mu}{}_{\nu},
\quad
{}^{(5)}G^{z}{}_{z}
=-\frac{6H_0^2}
       {\big(\sqrt{1+\beta^2}+\beta H_0 z\big)^2}.
\label{dS_int}
\eea


\section{Small thickness approximations}

Here we explain the small thickness approximation
which has been used in the text.
We use the representation 
$\epsilon(z) =\tanh\Big(\frac{z}{\sigma}\Big)$,
which leads to
$\delta_{\epsilon}(z)
=\frac{1}{2}\epsilon'(z)
=\frac{1}{2\sigma\cosh^2 \big(\frac{z}{\sigma}\big)}$.
In the limits of $\sigma\to 0$, it approaches
the usual sign 
$s(z)$ and delta functions
$\delta (z)$, respectively.
Thus $\sigma$
denotes the thickness of the codimension-2 brane.
Here we assume $\sigma H\ll 1$;
namely,
the brane thickness is much smaller than the
size of the cosmological horizon,
which is reasonable.
Keeping the leading order corrections due to a finite thickness,
for example,
the integral term
in the 
effective dark energy density, Eq. (\ref{effcomp})
becomes
\bea
&&\int_{-\infty}^{\infty} dz
\frac{\beta\delta_{\epsilon}(z)}
     {\sqrt{1+\beta^2(1-\epsilon(z)^2)}}
\Big[
1+\frac{\beta\epsilon(z) z}{\sqrt{1+\beta^2}}
\Big(
H+\frac{\dot{H}}{H}
\Big)
\Big]^2
=\frac{1}{2}
\int_{-1}^{1} 
\frac{\beta d\epsilon}
     {\sqrt{1+\beta^2(1-\epsilon^2)}}
\Big[
1+\frac{\beta\epsilon z(\epsilon)}{\sqrt{1+\beta^2}}
\Big(
H+\frac{\dot{H}}{H}
\Big)
\Big]^2
\nonumber\\
&\simeq& 
\frac{1}{2}
\int_{-1}^{1} 
\frac{\beta d\epsilon}
     {\sqrt{1+\beta^2(1-\epsilon^2)}}
\Big[
1+\frac{2\beta\sigma\epsilon^2}{\sqrt{1+\beta^2}}
\Big(
H+\frac{\dot{H}}{H}
\Big)
\Big]
\simeq \arctan (\beta)
+
\Big(
H+\frac{\dot{H}}{H}
\Big)
\sigma
C(\beta),
\eea
where $C(\beta)$ is defined in Eq. (\ref{cbet})
and shown in Fig. 2.
The same procedure is also applied for the 
integral term in the effective pressure Eq. (\ref{effcomp}).\\




\begin{thebibliography}{99}


\bibitem{6}G. Dvali, S. Hofmann and J. Khoury, Phys. Rev. D {\bf 76}
084006 (2007).




\bibitem{7}C. de Rham, S. Hofmann, J. Khoury and A. J. Tolley, JCAP 0802 (2008)011.

\bibitem{7a}C. de Rham, Can. J. Phys.{\bf 87} 201(2009)


\bibitem{gfree2}
  C.~de Rham, G.~Dvali, S.~Hofmann, J.~Khoury, O.~Pujolas, M.~Redi and A.~J.~Tolley,
  Phys.\ Rev.\ Lett.\  {\bf 100}, 251603 (2008)
  [arXiv:0711.2072 [hep-th]].




\bibitem{degra} 
  N.~Arkani-Hamed, S.~Dimopoulos, G.~Dvali and G.~Gabadadze,
  hep-th/0209227.


\bibitem{3}G. R. Dvali, G. Gabadadze and M. Porrati, Phys. Lett. B 485
(2000) 208.




\bibitem{gfree}
  C.~de Rham, J.~Khoury and A.~J.~Tolley,
  Phys.\ Rev.\  D {\bf 81}, 124027 (2010)
  [arXiv:1002.1075 [hep-th]].



\bibitem{cosmo}
  N.~Agarwal, R.~Bean, J.~Khoury and M.~Trodden,
  Phys.\ Rev.\  D {\bf 81}, 084020 (2010)
  [arXiv:0912.3798 [hep-th]].


\bibitem{gal}
  N.~Chow and J.~Khoury,
  Phys.\ Rev.\  D {\bf 80}, 024037 (2009)
  [arXiv:0905.1325 [hep-th]].

\bibitem{aga}
  N.~Agarwal, R.~Bean, J.~Khoury and M.~Trodden,
  arXiv:1102.5091 [hep-th].


\bibitem{nb}
  J.~Khoury and M.~Wyman,
  Phys.\ Rev.\  D {\bf 80}, 064023 (2009)
  [arXiv:0903.1292 [astro-ph.CO]].


\bibitem{mk}
  M.~Wyman and J.~Khoury,
  Phys.\ Rev.\  D {\bf 82}, 044032 (2010)
  [arXiv:1004.2046 [astro-ph.CO]].




\bibitem{4}
C. Deffayet, Phys. Lett. B
502 (2001) 199. 


 \bibitem{5}C. Deffayet, G. R. Dvali
and G. Gabadadze, Phys. Rev. D 65 (2002) 044023.


\bibitem{kalo} 
  N.~Kaloper,
  JHEP {\bf 0405}, 061 (2004)
  [hep-th/0403208].

\bibitem{ckt} 
  O.~Corradini, K.~Koyama and G.~Tasinato,
  Phys.\ Rev.\ D {\bf 77}, 084006 (2008)
  [arXiv:0712.0385 [hep-th]].

\bibitem{ckt2} 
  O.~Corradini, K.~Koyama and G.~Tasinato,
  Phys.\ Rev.\ D {\bf 78}, 124002 (2008)
  [arXiv:0803.1850 [hep-th]].



\end{thebibliography}
\end{document}